\documentclass{elsart3}

 \usepackage{graphicx}

\usepackage{amssymb}

\begin{document}

\begin{frontmatter}



\title{Electron transport through a strongly correlated monoatomic chain}


\author{M. Krawiec \corauthref{cor1}}
\corauth[cor1]{Corresponding author, tel.: +48 81 537 6146, 
               fax: +48 81 537 6191}
\ead{krawiec@kft.umcs.lublin.pl}
and
\author{T. Kwapi\'nski}
\ead{tomasz.kwapinski@umcs.lublin.pl}

\address{Institute of Physics and Nanotechnology Center, M. Curie-Sk\l odowska 
University, \\
pl. M. Curie Sk\l odowskiej 1, 20-031 Lublin, POLAND}

\begin{abstract}

We study transport properties of a strongly correlated monoatomic chain coupled 
to metallic leads. Our system is described by tight binding Hubbard-like model 
in the limit of strong on-site electron-electron interactions in the wire. 
The equation of motion technique in the slave boson representation has been 
applied to obtain analytical and numerical results. Calculated linear 
conductance of the system shows oscillatory behavior as a function of the wire 
length. We have also found similar oscillations of the electron charge in the 
system. Moreover our results show spontaneous spin polarization in the wire. 
Finally, we compare our results with those for non-interacting chain and 
discuss their modifications due to the Coulomb interactions in the system. 

\end{abstract}


\end{frontmatter}


\section{Introduction}
\label{intro}

Recently one-dimensional (1D) quantum wires (QW) have attracted much attention 
due to their potential applications in nanoelectronics \cite{Bowler} and 
quantum computing \cite{Bertoni}. The knowledge of the transport properties of 
such structures is crucial for the design and fabrication of the nanodevices. 
On the other hand the quantum wires, although conceptually simple, are very 
interesting from scientific point of view as they display extremely rich 
phenomena, very often different from those in two and three dimensions 
\cite{Luttinger,Haldane}. The understanding of the properties of such 1D 
objects is a major challenge in the field of nanophysics. 

The conductance of the quantum wires has been studied both experimentally and
theoretically by number of authors (see \cite{Agrait} for a review). The 
experimental studies require advanced techniques of fabrication of such 
structures. Those include: growing of QW on metallic surfaces
\cite{Jalochowski,MK_1}, scanning tunneling microscope techniques 
\cite{Yazdani} or mechanically controlled break junctions 
\cite{Agrait,Muller,Smit}. Those fabrication techniques allowed for revealing 
of many phenomena like charge quantization in units of $G_0 = 2e^2/h$
\cite{Wees}, deviations from that ($0.7 (2e^2/h)$ anomaly) \cite{Thomas},
spin-charge separation (Luttinger liquid) \cite{Auslaender}, oscillations of
the conductance as a function of the length of the chain \cite{Muller,Smit} or
spontaneous spin polarization in QW \cite{Thomas,Kane}.

The purpose of the present paper is two fold. The first one is to investigate 
the oscillations of the conductance as a function of the wire length in the 
case of strong Coulomb interactions. The oscillatory behavior of the 
conductance manifests itself as a maximum of the conductance when a number of 
the atoms in a wire is odd and minimum when the number is even. This effect is 
known as the even-odd conductance oscillations. Most common examples are the 
oscillations with a period of two 
\cite{Muller,Smit,Sim,Lang,Emberly,Gutierrez} and four atoms \cite{Thygesen}. 
However, the conductance can oscillate with different (from two and four) 
periods, depending on the average occupation of the wire. Moreover, recently 
the analytical formulas for M-atom (M $\geq$ 2) oscillations have been found 
\cite{TK_1}. However, those analytical formulas remain valid for noninteracting 
wire only. In the presence of strong Coulomb interactions the even-odd 
oscillations with a period of two atoms have been also found 
\cite{Oguri_1,Tanaka,Oguri_2,Oguri_3,Molina_1,Meden}. The M $\geq$ 2 
conductance oscillations have been only reported for the nearest neighbor 
Coulomb interactions \cite{Molina_2}. Therefore we shall study the oscillations 
of the conductance in the case of strong on-site Coulomb interactions and see 
how their period will be modified. 

The second purpose is to see if the wire will exhibit any spontaneous spin 
polarization in the presence of strong correlations. Such spontaneous 
polarization has been observed experimentally \cite{Thomas,Kane}. But it is 
well known that the ferromagnetism in strictly 1D objects is forbidden due to 
the Lieb-Mattis theorem \cite{Lieb}.

The paper is organized as follows. In Sec. \ref{model} we present theoretical 
description of the model wire, in Sec. \ref{results} we show the results of the 
calculations of the conductance, charge and spin polarization. Finally, in Sec. 
\ref{conclusions} we provide some conclusions. 


\section{Theoretical description}
\label{model}

Our system consists of the quantum wire modeled as a chain of $N$ atoms 
coupled to the left $L$ and right $R$ lead described by the following 
Hamiltonian in the limit of strong on-site Coulomb interaction 
($U_i \rightarrow \infty$) in the slave boson representation where the real 
wire electron $d_{i\sigma}$ is replaced by the product of the boson $b_i$ and
the fermion $f_{i\sigma}$ operators ($d_{i\sigma} = b^+_i f_{i\sigma}$)
\cite{Coleman,LeGuillou,MK_2}:
\begin{eqnarray}
H = \sum_{\lambda {\bf k} \sigma} \epsilon_{\lambda {\bf k}} 
c^+_{\lambda {\bf k} \sigma} c_{\lambda {\bf k} \sigma} +
\sum_{i\sigma} \varepsilon_i f^+_{i\sigma} f_{i\sigma}
\nonumber \\
+ \sum_{\langle i j \rangle, \sigma} 
t_{ij} f^+_{i\sigma} b_i b^+_j f_{j \sigma}
\nonumber \\
+ \sum_{{\bf k} \in L(R), \sigma} V_{L(R) {\bf k}} 
c^+_{L(R) {\bf k} \sigma} b^+_{1(N)} f_{1(N) \sigma} + h.c. ,
\label{Hamilt}
\end{eqnarray}
where $c_{\lambda {\bf k} \sigma}$ stands for the electron with the single
particle energy $\epsilon_{\lambda {\bf k}}$, the wave vector ${\bf k}$ and the
spin $\sigma$ in the lead $\lambda = L, R$. $\varepsilon_i$ denotes the wire 
energy level at site $i$, $t_{ij}$ is the hopping integral of the electrons 
between neighboring wire sites $i$ and $j$, and $V_{L(R) {\bf k}}$ is the
hybridization matrix element between electrons at site $1(N)$ and those
in the lead $L(R)$. 

In the linear response and at the zero temperature the conductance $G$ is
proportional to the total transmittance $T$, i. e. $G = \frac{2e^2}{h} T$. In 
our case the transmittance is given by \cite{TK_1,Datta}:
\begin{eqnarray}
T_N(E) = \sum_{\sigma} \Gamma_L \Gamma_R |G^r_{1N\sigma}(E)|^2 ,
\label{transmit}
\end{eqnarray}
where $G^r_{1N\sigma}$ is the retarded Green function (GF) connecting the ends 
of the wire (sites $1$ and $N$) and $\Gamma_{L(R)}$ is the elastic rate 
$\Gamma_{L(R)} = 2 \pi \sum_{\bf k} 
|V_{L(R) {\bf k}}|^2 \delta(E - \epsilon_{L(R) {\bf k}})$. In calculations we 
have assumed constant bare density of states in the leads. 

Using the equation of motion technique for the retarded GF with Hamiltonian 
(\ref{Hamilt}) one can write the general matrix equation for $G^r_{ij\sigma}$ 
in the form: 
\begin{eqnarray}
\hat A_{\sigma} \hat G^r_{\sigma} = \hat N_{\sigma} .
\label{EOM}
\end{eqnarray}
Due to the strong on site Coulomb interactions in the wire $U_i$, which is 
assumed to be infinity in our case, the problem cannot be solved exactly and 
one has to make approximations of the higher order GFs emerging in the equation 
of motion for the retarded GF 
$G^r_{ij\sigma} = \langle\langle b^+_1 f_{i\sigma} | 
f^+_{j\sigma} b_j \rangle\rangle_E$. We have used Hubbard I like approximation 
\cite{MK_2} according to which the GF 
$\langle\langle f^+_{i-\sigma} f_{i-\sigma} b^+_k f_{k\sigma} | 
f^+_{j\sigma} b_j \rangle\rangle_E$ is approximated by 
$\langle f^+_{i-\sigma} f_{i-\sigma} \rangle \langle\langle b^+_k f_{k\sigma} | 
f^+_{j\sigma} b_j \rangle\rangle_E$ and the other higher order GFs are
neglected. This approximation is reasonably good for not very large values of 
the hopping $t$ and neglects higher order processes, like for example the Kondo 
effect. 

Within the present approximation scheme $\hat A_{\sigma}$ in the Eq.(\ref{EOM}) 
is $N \times N$ tridiagonal symmetric matrix with the elements:
\begin{eqnarray} 
\hat A_{\sigma} = (E - \varepsilon_i) \delta_{i,j} + 
i \frac{\Gamma}{2} (\delta_{i,1} \delta_{j,1} + \delta_{i,N} \delta_{j,N}) -
\nonumber \\
t [(1 - n_{i-\sigma}) \delta_{i,i+1} + (1 - n_{i+1-\sigma}) \delta_{i+1,i}]
\label{Asig}
\end{eqnarray}
and $\hat N_{\sigma}$ is the diagonal matrix of the form:
\begin{eqnarray}
\hat N_{\sigma} = (1 - n_{i-\sigma}) \delta_{ij} ,
\label{Nsig}
\end{eqnarray}
with $n_{i\sigma} = \langle f^+_{i\sigma} f_{i\sigma} \rangle = 
-\frac{1}{\pi} \int dE Im G^r_{ii\sigma}(E)$ being the average occupation of 
the electrons with spin $\sigma$ at site $i$.


\section{Results}
\label{results}

In numerical calculations we have assumed all the wire site energies to be 
equal ($\varepsilon_i = \varepsilon_0$) and similarly hopping integrals 
$t_{ij} = t$. All energies are measured with respect to the leads Fermi energy 
$E_F = 0$ in units of $\Gamma = \Gamma_L = \Gamma_R = 1$. Moreover, the 
occupation $n_{i \sigma}$ is calculated self-consistently on each wire site.

To find the condition for M-atom conductance oscillations one has to solve the
relation: $T_N = T_{N+M}$, where $T_N$ ($T_{N+M}$) is the transmittance of the 
wire consisted of $N$ ($N+M$) atoms, given by Eq. (\ref{transmit}). In general,
for $U_i \rightarrow \infty$ it is not possible to get analytical expression 
for the oscillations condition without further assumptions. Note that for 
$U_i = 0$ the problem can be solved exactly and such condition can be found
\cite{TK_1}. In this case it reads $cos{\left(\frac{\pi l}{M}\right)} = 
\frac{E_F - \varepsilon_0}{2 t}$, where $l = 1$, $2$, ... $M-1$. 

In the case of $U_i \rightarrow \infty$ both matrices $\hat A_{\sigma}$ and 
$\hat N_{\sigma}$ depend on the occupation on each wire site $n_{i, \sigma}$ 
thus the problem has to be solved numerically. However, if one assumes that the 
occupation is the same on each wire site and does not depend on spin, i. e. 
$n_{i\sigma} = n_{\sigma} = n_{-\sigma}$, similar M-atom oscillations condition 
as for $U_i = 0$ can be found. In this case it reads:
\begin{eqnarray}
cos{\left(\frac{\pi l}{M}\right)} = 
\frac{E_F - \varepsilon_0}{2 t (1 - n_{\sigma})} ,
\label{cond_osc}
\end{eqnarray}
with $l = 1$, $2$, ... $M-1$. Unfortunately, one has to know the average
occupation $n_{\sigma}$. The only case of $M = 2$ can be solved analytically. 
As one can see from Eq. (\ref{cond_osc}) the period of two can be obtained for 
$E_F - \varepsilon_0 = 0$, i.e. when the wire single particle energy levels all 
coincide with the Fermi energy.

\begin{figure}[h]
 \resizebox{\linewidth}{!}{
  \includegraphics{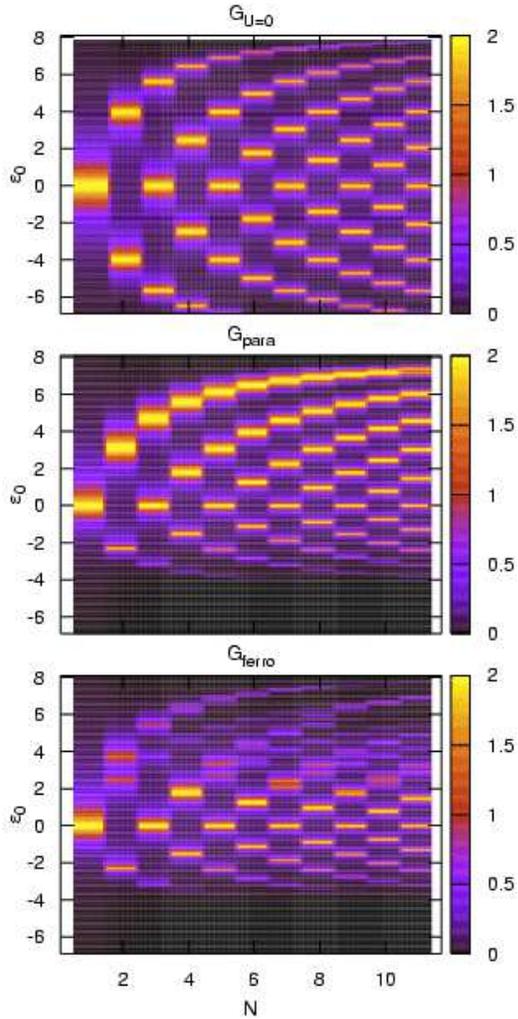}}
 \caption{\label{Fig1} The total linear conductance 
          $G = \sum_{\sigma} G_{\sigma}$ as a function of the wire length $N$ 
	  and the wire energy level $\varepsilon_0$ for $U_i = 0$ (top panel), 
	  $U_i \rightarrow \infty$ in paramagnetic configuration 
	  $n_{i\sigma} = n_{i-\sigma}$ (middle panel) and ferromagnetic one
	  (bottom panel).}
\end{figure}

In Fig. \ref{Fig1} the total linear conductance $G = \sum_{\sigma} G_{\sigma}$ 
is plotted as a function of the wire length $N$ and the energy level 
$\varepsilon_0$ for $t = 4$. 
As one can see all the figures show similar patterns with the regions of large
and small conductances. Moreover, for $\varepsilon_0 = 0$ behavior of the 
conductance is the same in all three cases which leads to the conclusion that
correlations are not important in this case. It always shows even-odd ($M = 2$) 
oscillations. Away from $\varepsilon_0 = 0$ correlations strongly modify the 
conductance, shifting the maxima of $G$ (for fixed $N$) towards lower (higher) 
energies for $\varepsilon_0 > 0$ ($\varepsilon_0 < 0$). This is due to the 
modification of the wire hopping, which depends now on the occupation 
$n_{i\sigma}$ (see Eq. (\ref{Asig})). This effect also leads to the strong 
asymmetry for positive and negative energies. For negative (positive) wire 
energies the occupation is large (small), therefore effective hopping 
$\tilde t = t (1 - n_{i-\sigma})$ is small (large), thus the conductance 
decreases (increases). Note that there is no such asymmetry in the case of 
$U_i = 0$. Moreover, different periods of the conductance oscillations can be 
observed, depending on the position of the wire energy level.

Another important finding is that the wire shows spontaneous spin polarization. 
It is well known that in strictly $1D$ wire the spin polarization is prohibited
due to the Lieb-Mattis theorem \cite{Lieb}. However this theorem is valid for
infinite wire only. In experimental situation the wire is always connected to 
the electrodes and this is why the spin polarization is observed experimentally
\cite{Thomas,Kane}. Interestingly, it was predicted recently that even infinite 
wire, but of the zig-zag shape, can also exhibit the spin polarization 
\cite{Klironomos}.

\begin{figure}[h]
 \resizebox{\linewidth}{!}{
  \includegraphics{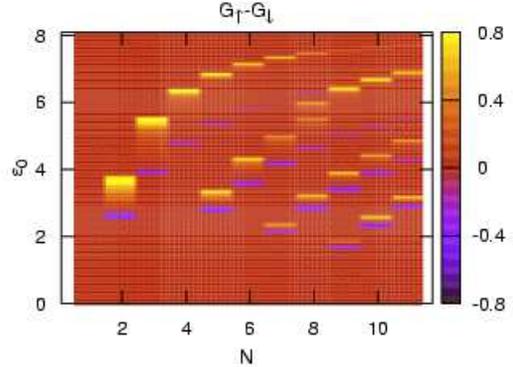}}
 \caption{\label{Fig2} The difference between spin up and spin down conductance
          as a function of the wire length $N$ and the wire energy level 
	  $\varepsilon_0$.}
\end{figure}

The total conductance in ferromagnetic case is shown in the bottom panel of 
Fig. \ref{Fig1}. Again, unlike for $U_i = 0$, the conductance pattern shows 
strong asymmetry for positive and negative energies. Note that for negative 
energies there are no differences between $G_{ferro}$ (bottom panel) and 
$G_{para}$ (middle panel). For such energies iterations always converge to 
paramagnetic solution. This can be again explained by effect of the hopping 
modification. For negative energies the wire occupation is large, thus the 
effective hopping $\tilde t = t (1 - n_{i-\sigma})$ is small. One can imagine 
in this case that the electrons are more localized and it is more convenient 
for them to spend more time on the same site than move to another one. 
Moreover in the case of the lack of the inter-site interactions any collective 
phenomenon is not possible. Situation is different for positive energies. In 
this case the electrons are more mobile, as the effective hopping is larger due 
to the small values of the wire occupations. Thus they interact with each other 
via hopping and it is possible and energetically more favorable to get the 
ferromagnetic state. The hopping, which is the correction to the position of 
the wire energy levels, leads to the effective splitting of this levels. Thus 
the conductance is spin polarized in this case. It can be read off from Fig. 
\ref{Fig2}, where the difference between spin up and spin down conductance is
displayed. 
It turns out that the strongest differences between $G_{\uparrow}$ and 
$G_{\downarrow}$ can be found for intermediate values of $\varepsilon_0$. For 
energies close to $E_F$ the modifications are weak due to the small values of 
the hopping while for higher energies the wire occupation is very small thus 
the difference between spin dependent effective hopping can be neglected. 
Finally it is worthwhile to note that for $N = 1$ (single atom) there is no 
ferromagnetic solution as the ferromagnetism is governed by inter-site hopping 
in our case. Whether this spontaneous spin polarization is a true effect or a 
drawback of the approximation used remains an open question, as it is known 
that the mean field like theories overestimate the role of the magnetism. The
problem will be further studied.

\begin{figure}[h]
 \resizebox{\linewidth}{!}{
  \includegraphics{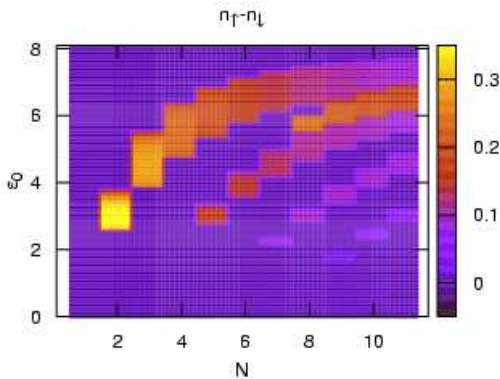}}
 \caption{\label{Fig3} The spin polarization as a function of the wire length 
          $N$ and the wire energy level $\varepsilon_0$.}
\end{figure}

Corresponding spin polarization $n_{\uparrow} - n_{\downarrow}$ 
($n_{\sigma} = \sum_i n_{i \sigma}/N$) is shown in Fig. \ref{Fig3}.
As one can see the spin polarization pattern is similar to that of the 
conductance differences (see Fig. \ref{Fig2}). 

\begin{figure}[h]
 \resizebox{\linewidth}{!}{
  \includegraphics{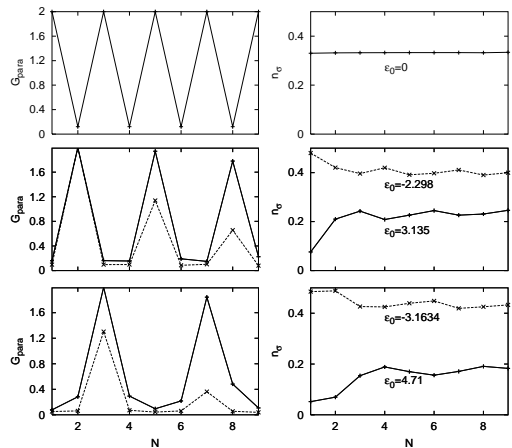}}
 \caption{\label{Fig4} The total conductance (left panels) and the occupation 
          (right panels) vs. wire length. The left panels show the conductance 
	  with different oscillations periods ($2$ to $4$) from top to bottom. 
	  The positions of the wire energy levels are indicated in the figure.}
\end{figure}

In Fig. \ref{Fig4} we show the conductance (left panels) and the occupation 
(right panels) as a function of the wire length for a number of the energy 
levels $\varepsilon_0$ in the paramagnetic configuration.
The values of $\varepsilon_0$ have been chosen in such a way that they lead to 
the maxima of the conductance for $N = 1$, $2$ and $3$ atom wire. For example, 
the maxima of the conductance of the $N = 2$ atom wire correspond to 
$\varepsilon_0 = -2.298$ and $3.135$ (see the middle panel of Fig. \ref{Fig1}).
These maxima in $U_i =0$ case give the periods of the conductance oscillations. 
To find $M$-atom period it is enough to determine the maxima of the conductance 
for $N = M - 1$ atom wire \cite{TK_1}. As one can read off from Fig. 
\ref{Fig4}, depending on $\varepsilon_0$, one gets different periods of the 
conductance oscillations. Moreover, except for the special case of $M = 2$, the 
amplitude of the oscillations decreases with the wire length. This is a kind of
damped oscillations. No such effect has been observed for $U_i = 0$ 
\cite{TK_1}. 

At this point we would like to comment on the other results known in the 
literature. The even-odd ($M = 2$) oscillations problem was extensively studied 
within the second order perturbation theory in $U_i$ (SOPT) 
\cite{Oguri_1,Tanaka} and the numerical renormalization group (NRG) approach 
\cite{Oguri_2,Oguri_3}. The results show similar behavior of the conductance 
for odd number of atoms in a wire: it always reaches the unitary limit 
($\frac{2 e^2}{h}$), independently of $U_i$. Such behavior is a consequence of 
the Kondo effect. However, in our case the situation is slightly different, as 
we get $M = 2$ oscillations in the mixed valence regime only 
($\varepsilon_0 = 0$), where the Kondo effect is excluded. Thus our even-odd 
oscillations are caused by the resonances associated with the energy level 
structure of the chain rather than the Kondo effect. On the other hand, the 
conductance for even number of atoms in a wire is strongly suppressed, in 
agreement with SOPT \cite{Oguri_1,Tanaka} and NRG approaches 
\cite{Oguri_2,Oguri_3}. However, NRG calculations show that the conductance 
exponentially depends on $U_i$, and in the limit of $U_i \rightarrow \infty$ 
vanishes, contrary to our results, as we get non-zero values of $G$ (see Fig. 
\ref{Fig4}), depending on the hopping $t$. The conductance vanishes in the 
limit of very large or very small values of $t$. Interestingly, when 
$t = \Gamma/2$, the conductance reaches the unitary limit and shows no
oscillations, i.e. is equal to $\frac{2 e^2}{h}$ for even and odd $N$.

Figure \ref{Fig4} (right panels) shows the wire length dependent occupation 
which also oscillates with the same period as the conductance does, except for 
the special case of $\varepsilon_0 = 0$ ($M = 2$), where the occupation remains 
constant. Moreover, the Coulomb interactions $U_i$ lead to the reduction of the 
occupation oscillation amplitude. Similar effect, albeit for small $U_i$, has 
been found within self-consistent Hartree-Fock approximation \cite{Kostyrko}. 

\begin{figure}[h]
 \resizebox{\linewidth}{!}{
  \includegraphics{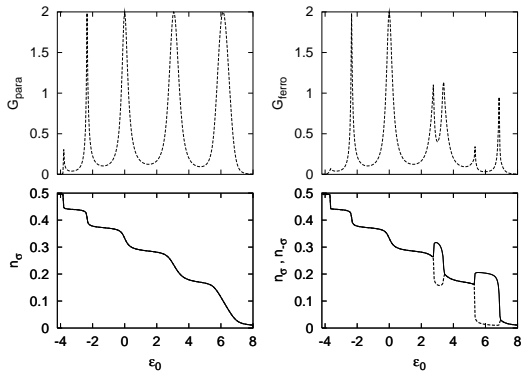}}
 \caption{\label{Fig5} The total conductance in paramagnetic (left top panel) 
          and in ferromagnetic (right top panel) configuration and 
	  corresponding occupations (bottom panels) vs. $\varepsilon_0$ of five 
	  atom wire.}
\end{figure}

In ferromagnetic case the situation is more complex. For positive values of 
$\varepsilon_0$ due to the splitting of the conductance maxima (see Fig.
\ref{Fig1}) no regular oscillations have been observed. On the other hand, for 
$\varepsilon_0 < 0$ one gets such oscillations but in this case the solutions 
remain always paramagnetic. Figure \ref{Fig5} shows the comparison of the 
conductance and the occupations in paramagnetic and ferromagnetic 
configurations for the wire consisted of five atoms.
For negative energies, as discussed before, there are paramagnetic solutions 
only. For $N > 1$ and $\varepsilon_0 > 0$ the ferromagnetic solutions emerge 
in certain energy regimes. In this case $n_{\uparrow} \neq n_{\downarrow}$ and 
resulting conductance peaks are split. Interestingly, the spin polarizations 
occurs only in the regimes where the occupation has a large slope or the total 
conductance has a maximum. Moreover, the splitting of the conductance leads to 
the fact that $G_{ferro}$ shows more maxima than $G_{para}$. A number of maxima 
of $G_{ferro}$ is related to the wire length (number of atoms - $N$) and for 
odd $N$ it gives $(3N-1)/2$ maxima, while for even $N$ one observes $3N/2$ 
maxima. In paramagnetic case it is always equal to a number of atoms $N$ in the 
wire (compare top panels of Fig. \ref{Fig5}).

Finally we would like to comment on the validity of our approach. The present
calculations completely neglect the Kondo effect which is important at low
temperatures and $\varepsilon_d < 0$. We expect some modifications in this
regime, as this effect leads to the corrections of the conductance of the order
of $\frac{e^2}{h}$. Thus our results apply for temperatures higher than the
Kondo temperature. On the other hand, we do not expect any qualitative 
modifications in the mixed valence and the empty regimes 
($\varepsilon_d \geq 0$).


\section{Conclusions}
\label{conclusions}

In summary we have studied the conductance oscillations of the strongly
interacting wire as a function of the wire length. We have found that strong
Coulomb interactions significantly modify the periods of the oscillations 
showing strong asymmetry for negative and positive wire energy levels. They 
also lead to the suppression of the conductance with increasing wire length.
There are no such effects for noninteracting wire. Moreover, strong 
interactions lead to the spontaneous spin polarization for positive wire 
energies, observed in experiments. 


\noindent {\bf Acknowledgements} \\
This work has been supported by the grant no.1 P03B 004 28 of the Polish 
Committee of Scientific Research. T. K. thanks the Foundation for Polish 
Science for a Fellowship for Young Scientists.


\end{document}